# Traffic model of LTE using maximum flow algorithm with binary search technique.

Md. Zahurul Haque*, Md. Rafiqul Islam **

* Department of Computer Science and Engineering, ** Department of Electrical and Electronic Engineering
Manarat International University, Dhaka.
*Email:* * *jahurulhaque@manarat.ac.bd,* **rafiqulislam@manarat.ac.bd*

*Abstract—In* recent time a rapid increase in the number of smart devices and user applications have generated an intensity volume of data traffic from/to a cellular network. So the Long Term Evaluation (LTE) network is facing some issues difficulties of the base station and infrastructure in terms of upgrade and configuration because there is no concept of BSC (Base Station Controller) of 2G and RNC (Radio Network Controller) of 3G to control several BTS/NB. Only 4G (LTE) all the eNBs are interconnected for traffic flow from UE (user equipment) to core switch. Determination of capacity of a link of such a network is a challenging job since each node offers its own traffic and at the same time conveys traffic of other nodes. In this paper, we apply maximum flow algorithm including the binary search technique to solve the traffic flow of radio network and interconnected eNBs of the LTE network. The throughput of the LTE network shown graphically under the QPSK and 16-QAM.

## I. INTRODUCTION

In LTE network there is no concept of BSC (Base Station Controller) of 2G and RNC (Radio Network Controller) of 3G to control several BTS/NB. In 4G (LTE) all the eNBs are interconnected for traffic flow from UE (user equipment) to core switch. In this thesis the traffic flow of such network is modeled by a newly proposed algorithm of maximum flow (Ford-Fulkerson) with binary search technique.

Rapid increase of the number of smart devices and user applications has generated an intensity volume of data traffic from/to a cellular network. In [1] authors introduced a new network architecture for LTE and Wi-Fi slicing networks taking into account the advantage of software defined networking (SDN) and network function virtualization (NFV) concepts.

The current LTE network is facing some issues with the architecture in terms of centralized data flow bearers, centralized monitoring and control, and difficulties of base station and infrastructure in terms of upgrade and configuration is found in [2].

Using Wi-Fi hotspot is increasingly adopted by smartphones and operators as it boosts network capacity. In [3] authors proposed a handover decision algorithm from LTE-A to WMN is presented and aims to maximize the amount of offloaded data through the WMN as the last network uses an unlicensed spectrum.

The maximum flow problems involve finding a feasible flow through a single-source, single-sink flow network that is maximum. The Ford-Fulkerson method of [4] (named for L. R. Ford, Jr. and D. R. Fulkerson) provides an algorithm which computes the maximum flow in a flow network. The name "Ford-Fulkerson" is often also used for the Edmonds-Karp algorithm, which is a specialization of Ford-Fulkerson. The idea behind the algorithm is simple. As long as there is a path from the source (start node) to the sink (end node), with available capacity on all edges in the path, we send flow along one of these paths. Then we find another path, and so on. A path with available capacity is called an augmenting path.

The problem of finding a maximum flow is a classical problem with a wide variety of scientific and engineering applications, such as engineering, management science etc. In a real network, the maximum flow problem solves in [5] and showed how to transmit the maximal matter flow from one point to another point along the network links when the capacity requirement is met.

In [5-6] author proposed a new augmenting path based algorithm which is called draining algorithm to find the maximum flow from some specific source to destination. Unlike other augmenting path based algorithms the proposed algorithm drains the unnecessary capacities out of the graph network to achieve the maximum flow.

A genetic algorithm based approach to hard test generation for the maximum flow algorithm. To reduce the amount of human effort, authors suggested new search-based optimization techniques, such as genetic algorithm explained in [7].

An algorithm is proposed in [8] to find a path which has the maximum allowed flow rate for data, between source and destination in a network. Using Prim's technique author found a path in a network between source and destination. Firstly, the maximum allowed flow rate is calculated and take the maximum one out of all maximum flow rates.

Objectives of the thesis work is to implement the modified form of maximum flow algorithm where intermediate nodes involve in generation of new flow. The concept is applied in interconnected eNBs of LTE network.

## II. MAXIMUM FLOW ALGORITHM

In graph theory, a flow of a network is defined as a directed graph involving a source(S) and a sink (T) and several other nodes connected with edges. Each edge has an individual capacity which is the maximum limit of flow that edge could allow.

It is defined as the maximum amount of flow that the network would allow to flow from source to sink. Multiple algorithms exist in solving the maximum flow problem. Two major algorithms to solve these kind of problems are **Ford-Fulkerson algorithm** and **Dinic's Algorithm**.

### A. Ford-Fulkerson Algorithm

Ford-Fulkerson method or Ford-Fulkerson algorithm was developed by L. R. Ford, Jr. and D. R. Fulkerson in 1956.

Let G (V, E) be a graph, and for each edges from u to v, let c (u, v) be the capacity and f (u, v) be the flow from u to v. We need to find the maximum flow from the source node s to sink node t. After each step following conditions should follow.

- The flow along an edge can't exceed its capacity i.e $f(u, v) <= c(u, v)$.
- Net flow in the edges follows skew symmetry i.e. $f(u, v) = -f(v, u)$ where $f(u, v)$ is flow from node u to node v.
- For any non-source and non-sink node, the input flow is equal to output flow.

We define the residual network $G_f(V, E_f)$ to be the network with capacity $c_f(u, v) = c(u, v) - f(u, v)$ and no flow. If the graph has multiple sources and sinks, we can create a dummy source or sink and connect this dummy nodes to corresponding sources or sinks with infinite edge capacity. The pseudo code of ford-Fulkerson algorithm is given below.

> Inputs Given a network G = (V, E) with capacity c, a source node s, and sink node t.
> Output Compute a maximum flow f from s to t
> 1. f(u, v)←0 for all edges(u, v)
> 2. While there is a path p from s to t in $G_f$, such that $c_f(u, v) > 0$ for all edges $(u, v) \in p$:
>    1. Find $c_f(u, v) = \min\{ c_f(u, v) : (u, v) \in p \}$
>    2. $f = f + c_f(u, v)$
>    3. For each edges $(u, v) \in p$
>       1. $f(u, v) \leftarrow f(u, v) + c_f(p)$
>       2. $f(v, u) \leftarrow f(v, u) - c_f(p)$

The path in step 2 can be found with breadth-first search or depth-first search algorithm.

## B. Augmenting Path

An augmenting path is a simple path from source to sink which do not include any cycles and that pass only through positive weighted edges. A residual network graph indicates how much more flow is allowed in each edge in the network graph. If there are no augmenting paths possible from S to T, then the flow is maximum. The result i.e. the maximum flow will be the total flow out of source node which is also equal to total flow in to the sink node.

### a) Implementation
1. An augmenting path in residual graph can be found using DFS or BFS.
2. Updating residual graph includes following steps:
   - For every edge in the augmenting path, a value of minimum capacity in the path is subtracted from all the edges of that path.
   - An edge of equal amount is added to edges in reverse direction for every successive nodes in the augmenting path.

### b) Complexity
The runtime complexity of Ford-Fulkerson algorithm is O (f * E), where f is maximum flow and E is number of edges of the network.

## III. Traffic model of LTE

The aggregate offered traffic on sink node,

$$A = \sum_{i=1}^{n} A_i = \lambda_T t_h ;$$

Where $\lambda T \rightarrow$ arrival rate of RB in RB/ms

$th \rightarrow$ service time of a RB in ms

We know the duration of a RB is 0.5 ms and the number of symbols in 0.5ms is 7 under normal CP. In QPSK each symbol carries 2 bits therefore the number of bits per RB per subcarrier is 2×7=14 bits.

Now the bit rate/subcarrier =14/0.5 = 28Kbps = 0.028 Mbps.

The aggregate offered traffic, $A = \lambda T th = 0.5 th$

Let the node $n$ is anchored with the core switch hence node $n$ is considered are the sink of the maximum flow algorithm. We will run the algorithm taking node $i = 1, 2, 3\ldots n$-1 as the source. We can run the following algorithm to get the maximum flow.

## a) Maximum flow algorithm with binary search technique

The offered data of node i= 1, 2, 3…, n-1 is $C_i$ and the edge (i, j) have weight $W_{ij}$. Let Y is the sum of $C_i$ for i = 1, 2, 3… n-1.

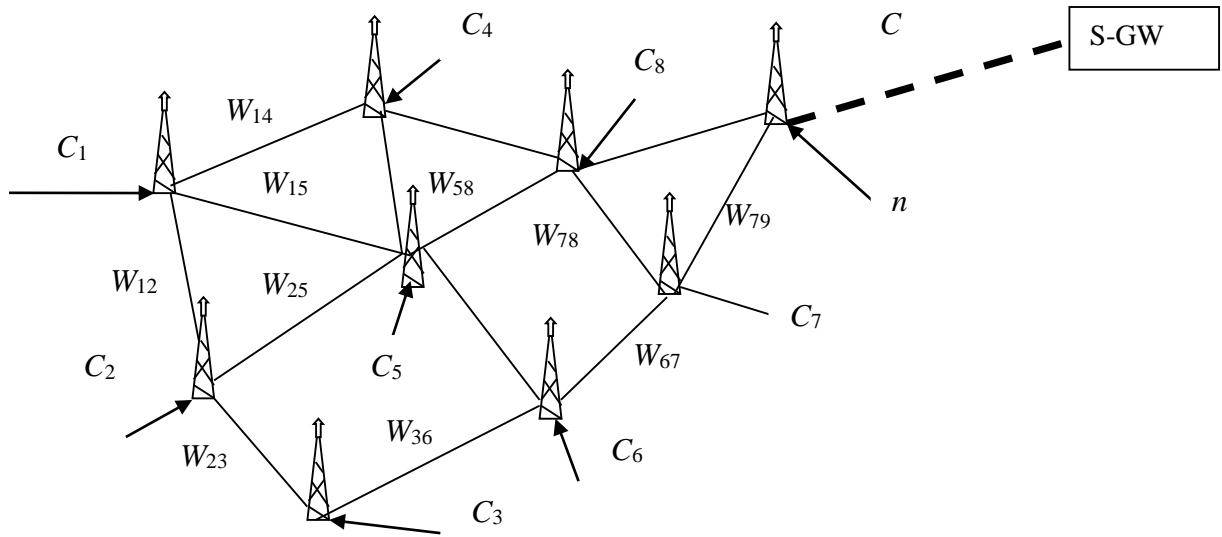

Figure 3.1 Interconnected eNBs of LTE

## b) Implementation

### Procedure

1. Set $L = 1$ Mbps, $R = Y$ Mbps.
2. If $L >= R$, go to step 9.
3. Set $M = (L + R)/2$ (the position of the middle element) to the floor, or the greatest integer less than $(L + R)/2$.
4. Determine the weighted graph of the connected eNBs of MME (Mobility Management Entity). Initialize all the edges (i, j) with the weight $W_{ij}$ between node i and j with M Mbps.
5. Take the dummy node 0 as a source node and add the edges (0, i) with edge weight $W_{0i} = C_i$.
6. Determine maximum flow **f** from node 0 to *n*.
7. If the maximum flow is equal to Y, then set $R = M - 1$ and go to step 2.
8. If the maximum flow is less than Y, then set $L = M + 1$ and go to step 2.
9. Hence the average cost at $L$ is optimal. So the run the maximum flow with edge capacity=L.

10. Now the edge capacity of c (i, j) will be the f (i, j).

*Pseudo code of Ford-Fulkerson algorithm*

**Inputs:** Given a network G = (V, E) with capacity M for each edges, a source node s, and sink node t.
**Output:** Compute a maximum flow f from s to t
1. f(i,j)←M  for all edges(i, j)
2. While there is a path p from s to t in Gf, such that cf(i, j) > 0 for all edges (i, j) ∈ p:
    1. Find cf(i, j) = min{ cf(i, j) : (i, j) ∈ p }
    2. f = f + cf(i, j)
    3. For each edges (i, j) ∈ p
        1. f(i, j)← f(i, j) + cf(p)
        2. f(i, j)← f(i, j)  - cf(p)
Returns total flow from s to t.

*Pseudo code for adjusting the edges capacity*

1. L← 1, R← Y
2. If L>= R go to step 7
3. M← (L + R)/2
4. Compute maximum flow f such that for each edges, capacity is M. [ 3.2.2 ]
5. If f >= Y, R ← M - 1 and go to step 2
6. If f < Y go to, L ← M + 1 and go to step 2
7. Compute flow for M = L
8. c(i, j)←f(i, j)  for all edges(i, j)

## d) Blocking Probability of the network

Let the total flow is found as *C* Mbps/subcarrier.
Since the bit rate/subcarrier of a RB under QPSK is 0.028 Mbps therefore the number of RB carried by the network simultaneously, $N = \left\lfloor \dfrac{C}{0.028} \right\rfloor$.

Let the field survey provides that each user on an average claims *m* RB per call. If the capacity of the network is *N* RB then the number of channel of the network will be, $k = \left\lfloor \dfrac{N}{m} \right\rfloor$.

Each user on the network generates *s* calls/min and *m* RB are booked on each call and the duration of a call is *th* min. The offered traffic per user in terms of RB is, $a = (s.m)th = smth$
The total offered traffic, $A = aM = smthM$
Example-1
Let *M*= 200, *m*=3 RB/call/user, *th* =1.5 mins and *s* = 1/60 min. If the current capacity of the network is 25 RB then find the offered traffic and number of channel.
$A = smthM = (1/60).3. (1.5) 200 = 15$ Erls.

The number of channels, $k = \left\lfloor \dfrac{N}{m} \right\rfloor = \left\lfloor \dfrac{25}{3} \right\rfloor = 8$

The blocking probability,

$$B(A,k) = \dfrac{\dfrac{A^k}{k!}}{\sum_{i=0}^{k} \dfrac{A^i}{i!}}$$

If $B(A, k)$ is weighted with the SER of QPSK and 16-QAM we get the relation of network traffic and SNR of the link between UE and eNB.

## IV. Results

In LTE two main types of modulation schemes are used: QPSK and 16-QAM. When users are adjacent to eNB can secure strong SNR then system use 16-QAM (4 bits/symbol) again when users are at cell boundary experience huge fading then use QPSK scheme (2bits/symbol). Therefore throughput is higher when users are closed to eNB. In section we first consider the case of QPSK scheme. The weighted graph of fig 4.1 is the interconnected eNB where each node corresponds of eNB and the edges are the wireless link among eNBs. Here we assume that each node offers the data of 2Mbps/subcarrier. The weight of each edge is the channel capacity (in Mbps) of that link. Here we use nine eNBs and found the maximum flow from $i$th node to node 9 (using Matlab 18) as shown in table I.

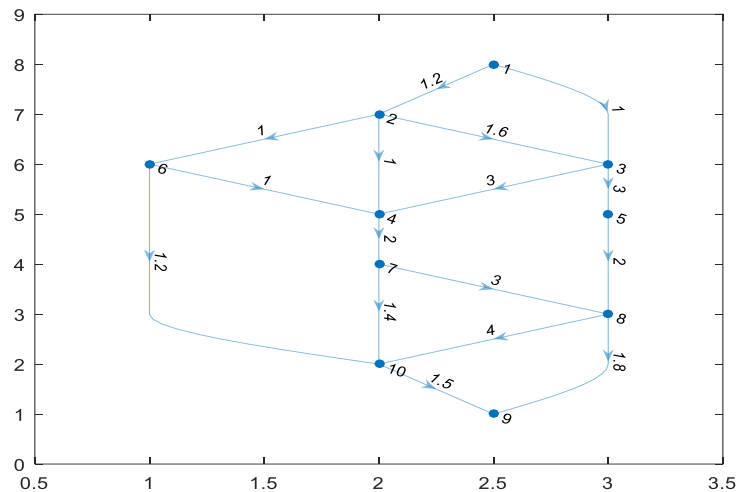

Figure 4.1 Equivalent weighted graph of interconnected eNBs of LTE

Table-I
Maximum flow per subcarrier

| Source | Sink | Maximum flow (Mbps/subcarrier) |
|---|---|---|
| 1 | 9 | 2.2 |
| 2 |   | 3.3 |
| 3 |   | 3.3 |
| 4 |   | 2 |
| 5 |   | 2 |
| 6 |   | 2.2 |
| 7 |   | 3.3 |
| 8 |   | 3.3 |

First of all we consider, the number of users of $M$ = 50,100 and 200, average RB /call of $m$ =5, arrival rate per user 1/60min and modulation scheme of QPSK. As a numerical example we consider a network of low capacity i.e. C = 2.2 Gbps. Fig 4.2 shows the variation of 'blocking probability' against 'offered average RB /call' taking 'average holding time' as a parameter. Here blocking probability increases with increase in 'offered average RB /call', number of users and 'holding time'. The rise of all curves are found exponential and mutually parallel. Actually 'offered average RB /call' $m$ is proportional to offered traffic at the same time inversely proportional of the number of channel hence fig.2 actually shows the curves equivalent to offered traffic vs. B of Erlang's model.

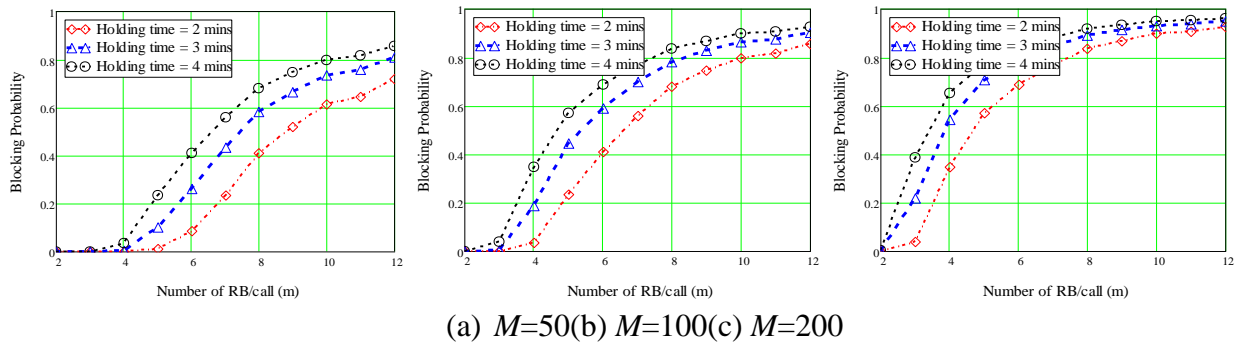

(a) $M$=50 (b) $M$=100 (c) $M$=200

Figure 4.2 Variation of offered RB/call ($m$) against blocking probability

The variation of blocking probability against the 'simultaneous capacity of RB' taking 'average holding time' as a parameter is shown in fig.3. Three graphs are shown for $m$ = 2, 3 and 4. Actually the 'simultaneous capacity of RB' is equivalent to the number of channels hence the curves of fig 4.3 is equivalent to the curve of 'number of channel' vs. blocking probability. The blocking probability decreases with increase in 'simultaneous capacity of RB' but reveres relation is found with 'average holding time'. The blocking probability also increases with increase in $m$ is also visualized from fig 4.3 (a)-(c).

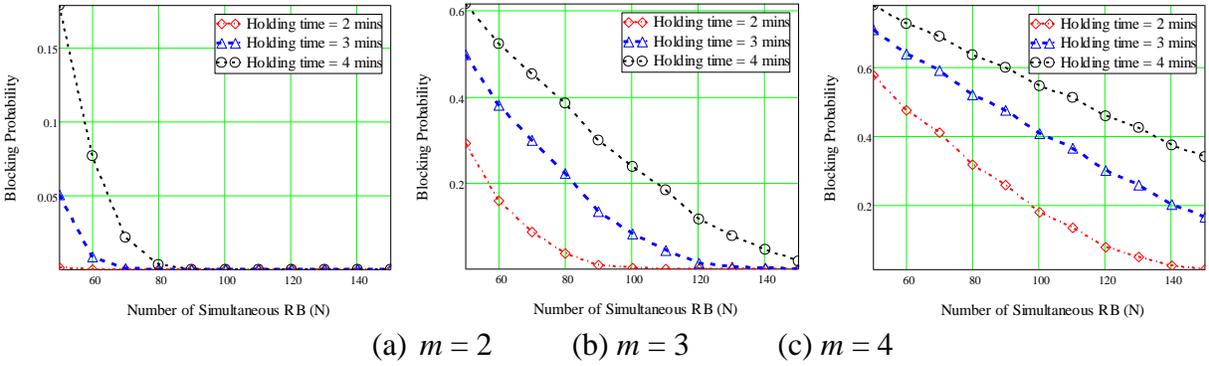

(a) *m* = 2     (b) *m* = 3     (c) *m* = 4

Figure 4.3 Variation of number of simultaneous RB (*N*) against blocking probability

Next we consider a network of higher capacity as shown in fig 4.4; where we assume that each node offers the data of 4 Mbps/subcarrier. For example using maximum flow algorithm we got the maximum flow from *i*th node to node 9 is shown in table II. Let the number of user is *M* = 50, 100 and 200 and the capacity C= 9 Gbps to observe the impact of capacity of source to sink like before. The blocking probability is found lower than the previous case shown in fig 4.5 (a)-(c).

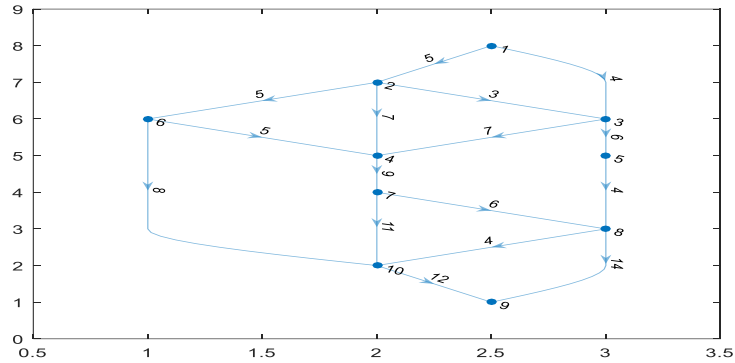

Figure 4.4 Equivalent weighted graph of interconnected eNBs of LTE of higher capacity

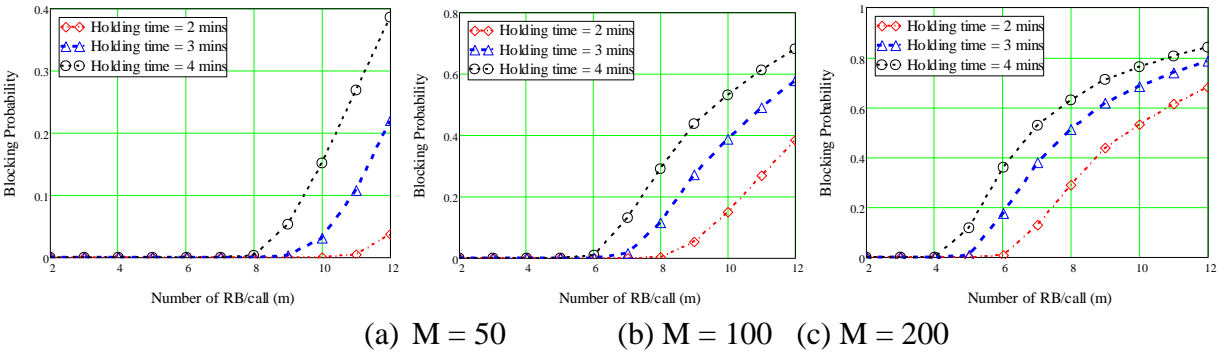

(a) M = 50     (b) M = 100     (c) M = 200

Figure 4.5 Variation of offered RB/call (*m*) against blocking probability under QPSK

Table-II
Maximum flow per subcarrier

| Source | Sink | Maximum flow (Mbps/subcarrier) |
|---|---|---|
| 1 | 9 | 9 |
| 2 |  | 15 |
| 3 |  | 11 |
| 4 |  | 9 |
| 5 |  | 4 |
| 6 |  | 13 |
| 7 |  | 17 |
| 8 |  | 18 |

In 16-QAM the number of simultaneous RB carried by the network becomes, $N = \left\lfloor \dfrac{C}{0.056} \right\rfloor$ i.e. half of QPSK. Therefore the performance is deteriorated under 16-QAM visualized from fig 4. 6. Actually $N$ is taken as the number of traffic channel of Erlang's model of previous section, is half compared to QPSK but throughput remains same since each symbol of 16-QAM carries twice the bits of QPSK. Therefore the RR/call i.e. $m$ under 16-QAM is just the half of QPSK to keep the same throughput hence actually performance remains same like fig 4.5. If we could relate the traffic model with SNR of link then we could get better performance under 16-QAM at higher SNR.

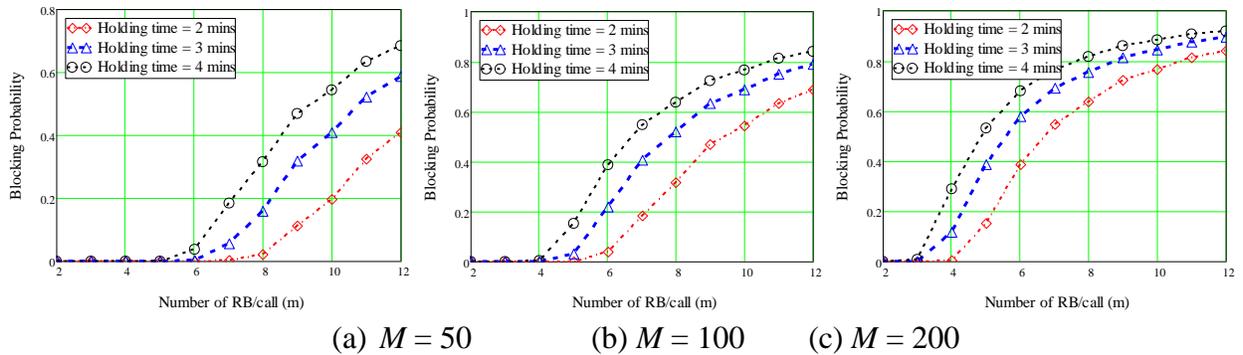

(a) $M = 50$    (b) $M = 100$    (c) $M = 200$

Figure 4.6 Variation of RB/call against $B$ under 16-QAM

## V. Conclusion

In this thesis work, we proposed a new model of Evolved UTRAN in which only one S1 link is used by the eNB nearest to S-GW. Since each eNB offers some traffic hence the flow generation rule of maximum flow algorithm is violated. Hence we applied a new proposed algorithm. The proposed algorithm can also be applied in an interconnected wired network of a metropolitan area like multiple telephone exchange connected to a tandem exchange. In order to fix an optimal capacity of each link of the interconnected eNB, we used maximum flow algorithm along with binary search technique. Due to binary search technique, every link is used with approximately equal probability hence any link won't be ever under immense traffic pressure. If any link is broken, an alternative path can be chosen in order to pass the traffic and traffic loss is minimized by keeping carried traffic smaller than offered traffic. But we used Ford-Fulkerson maximum flow algorithm with linear time complexity which can be reduced to logarithmic time using Dinic's algorithm in future.